# Mediation effects that emulate a target randomised trial: Simulation-based evaluation of ill-defined interventions on multiple mediators

Margarita Moreno-Betancur[1,2,*], Paul Moran[3], Denise Becker[2], George C Patton[1,2], John B Carlin[1,2]

[1]University of Melbourne, Melbourne, Australia
[2]Murdoch Children's Research Institute, Melbourne, Australia
[3]University of Bristol, Bristol, U.K.
[*] Corresponding author: margarita.moreno@mcri.edu.au

## Abstract

Many epidemiological questions concern potential interventions to alter the pathways presumed to mediate an association. For example, we consider a study that investigates the benefit of interventions in young adulthood for ameliorating the poorer mid-life psychosocial outcomes of adolescent self-harmers relative to their healthy peers. Two methodological challenges arise. Firstly, mediation methods have hitherto mostly focused on the elusive task of discovering pathways, rather than on the evaluation of mediator interventions. Secondly, the complexity of such questions is invariably such that there are no existing data on well-defined interventions (i.e. actual treatments, programs, etc.) capturing the populations, outcomes and time-spans of interest. Instead, researchers must rely on exposure (non-intervention) data to address these questions, such as self-reported substance use and employment. We address the resulting challenges by specifying a target trial addressing three policy-relevant questions, regarding the impacts of hypothetical (rather than actual) interventions that would shift the mediators' distributions (separately, jointly or sequentially) to user-specified distributions that can be emulated with the observed data. We then define novel interventional effects that map to this trial, emulating shifts by setting mediators to random draws from those distributions. We show that estimation using a g-computation method is possible under an expanded set of causal assumptions relative to inference with well-defined interventions. These expanded assumptions reflect the lower level of evidence that is inevitable with ill-defined interventions. Application to the self-harm example using data from the Victorian Adolescent Health Cohort Study illustrates the value of our proposal for informing the design and evaluation of actual interventions in the future.

# INTRODUCTION

In areas such as life course and social epidemiology, questions arise around potential interventions to alter pathways presumed to mediate an association, such as between an early-life marker of vulnerability and later outcomes. Our motivating example investigated potential interventions to counter the poorer psychosocial outcomes in adulthood of adolescents who self-harm relative to their healthy peers, such as targeting substance use and mental health problems in young adulthood. Addressing such questions raises two key methodological challenges, which this paper aims to tackle.

The first challenge relates to the focus of the mediation literature on the discovery of mechanistic pathways.[1] The prevailing logic is to assume a pre-existing (axiomatic) notion of mediation and then to define "indirect" effects so as to detect and quantify this, with the modern definitions in the potential outcomes framework referred to as "natural effects".[2–5] These effects are not defined in a way that makes them empirically measurable, even hypothetically, in a randomised experiment,[1,6] and alternative methods that would explicitly address the issue of mediator intervention evaluation have been lacking. This is striking given that the implied appeal of discovering pathways is often to reveal potential intervention points. It also contrasts with current thinking in the broader epidemiological literature, where the elusive nature of the notion of "causation" [7,8] (of which "mediation" is an extension), tied to aspirations for an epidemiology of consequence,[9] has brought a move away from the quest for the discovery of causes. Instead, emphasis is given to the more tangible goal of assessing effects of causes conceptualised as interventions,[7,10–12] with analyses designed to emulate a "target trial",[13,14] defined as the ideal randomised trial that one would hypothetically perform to evaluate the intervention in question.

The second challenge is that the endeavour of intervention evaluation presupposes well-defined interventions. However, the complexity of the questions being asked in many areas, such as the self-harm example, is often such that there are no existing data on well-defined interventions that also capture the populations, outcomes and time-spans of interest. Instead, researchers have to rely on observational exposure (non-intervention) data, for example from long-term longitudinal cohort studies, to address their questions. There has been much criticism of such "exposure epidemiology" for causal inference, yet producing some evidence, even if imperfect, is arguably a key first step to future intervention development and evaluation.[15] This explains a recent push for addressing, rather than shunning, the methodological challenge of ill-defined interventions.[10,15–17]

In this work, we reverse the logic that has driven the mediation literature: rather than assuming a pre-existing notion of mediation, we propose to start with specific policy-relevant questions relating to mediator interventions and then define effects to address these in explicit correspondence to a target trial. We show that, within this logic, mediation effects are not required if the question and available data pertain to well-defined mediator interventions, but mediation regains its relevance in the context of ill-defined interventions, in the form of so-called "interventional effects" (a.k.a. "interventional randomised analogues")[18–21].

Specifically, recent work shows that interventional mediation effects implicitly emulate effects defined by target trials that evaluate the impacts of distributional shifts in the mediators.[22] We propose that conceptualising such distributional shifts as arising from hypothetical interventions provides a useful framework for tackling the issue of ill-defined mediator interventions. In particular, this acknowledges the composite nature of the exposures under consideration[16] while being explicit about the additional assumptions required. However, given their unintentional (implicit) nature, the target trials emulated by estimands that have previously been considered under the "interventional effects" umbrella[18–21,23–28] are not necessarily relevant for informing policy. Therefore, we define novel interventional effects explicitly in terms of a target trial addressing three specific policy-relevant questions, regarding the impacts on intervening to shift mediators separately, jointly or sequentially. Importantly, our goal was to define the contrasts of interest in the context of these questions and ill-defined interventions, acknowledging that this is only the first step of a full "target trial approach", which must also consider further protocol components of the target trial.[14]

The paper is structured as follows. First, we introduce the self-harm example. Second, we introduce the issue of ill-defined mediator interventions and propose a set of principles for tackling it via simulation of hypothetical interventions. Third, we describe the target trial integrating these principles and derive novel definitions of interventional effects that map to that trial. Fourth, we determine identification assumptions and describe a g-computation estimation method, providing example R code. Finally, we illustrate the value of the proposed effects in the self-harm example and conclude with a discussion.

**SELF-HARM EXAMPLE**

Adolescent self-harm is on the rise [29–31] and is associated with substantial disease burden[32] through immediate effects on health and mortality,[33] as well as through persisting associations with poor health and social functioning in later life, including higher rates of substance use,[34,35] depression[34] and financial hardship.[36] A question of considerable public health interest is whether policies targeting young adulthood processes may have benefit in reducing these impacts. We focus on the financial hardship outcome, and consider four mediators: depression or anxiety, cannabis use, lack of higher education and unemployment.[36] We draw data from the Victorian Adolescent Health Cohort Study, a 10-wave longitudinal population-based cohort study of health across adolescence to the fourth decade of life in the state of Victoria, Australia (1992-2014). Data collection protocols for this study were approved by the Ethics in Human Research Committee of the Royal Children's Hospital, Melbourne. Informed parental consent was obtained before inclusion in the study. In the adult phase, all participants were informed of the study in writing and gave verbal consent before being interviewed. The Supplementary Materials provide more details on study design, with the key measures of relevance for our illustrative analysis summarised next.

The main exposure, denoted $A$, was adolescent self-reported self-harm across waves 3 to 6 (age 15–18 years), with $A = 1$ if self-harm was present at any wave during adolescence and $A = 0$ otherwise, including when all wave-specific measures were either negative or missing. The outcome ($Y$) was self-reported financial hardship at wave 10 (median age 35 years), with $Y = 1$ if financial hardship was present and $Y = 0$ otherwise. The mediators, measured at wave 8 (median age 24 years), were depression or anxiety ($M_1$), weekly or more frequent cannabis use over the past year ($M_2$), not having completed a university degree ($M_3$), and not being in paid work ($M_4$). We define $M_k = 1$ if the mediator was present and $M_k = 0$ if it was absent ($k = 1, ..., 4$).

Pre-exposure confounders ($C$) of the exposure-mediator, mediator-outcome and exposure-outcome associations were selected on an a priori basis: participant sex, parental completion of high school (as a marker of socio-economic position), parental divorce or separation up to and including wave 6, and adolescent antecedents of the mediators where present, specifically participant completion of high school, adolescent depression or anxiety and cannabis use (weekly or more frequent). The latter two were summarised across waves 3-6 in the same way as the exposure. Figure 1 shows the assumed causal structure for the observed data following prior evidence.[30,34–36] Although the mediators are assumed to be correlated, the causal diagram is agnostic to their causal ordering.

**TACKLING ILL-DEFINED INTERVENTIONS**

We consider the general case of $K$ mediators and, initially, the question of assessing the impact in the exposed ($A = 1$) of $K$ hypothetical interventions, each targeting a single mediator (in the next section we consider other possibilities). Let $B_k = 1$ if the intervention targeting $M_k$ is received and $B_k = 0$ if not ($k = 1, ..., K$). If these interventions were well-defined and existed, for instance, in the form of specific programs for mental health care, substance use reduction, and career development in the self-harm example, and we had relevant data, we could address the questions of interest by separately assessing the effect of each intervention in the exposed. That is, letting $Y_{B_k=b_k}$ denote the potential outcome when setting $B_k = b_k$, we would compute and compare the causal effects (in the difference scale) as $E(Y_{B_k=1}|A = 1) - E(Y_{B_k=0}|A = 1)$, $k = 1, ..., K$. Here the unexposed group and mediation effects are not relevant.

However, interventions are often ill-defined, as in the example. A common "simple" approach in that case would be to replace $B_k$ by $M_k$ and estimate the causal effects $E(Y_{M_k=1}|A = 1) - E(Y_{M_k=0}|A = 1)$, but this raises the following issues. The potential outcomes $Y_{M_k=m_k}$ are ill-defined: for example, there are many potential interventions for improving the mental health of individuals that could lead to very different conclusions regarding causal effects.[37,38] Furthermore, any such intervention is unlikely to result in complete elimination of depression and anxiety in the self-harm group, which is the scenario that $E(Y_{M_k=0}|A = 1)$ corresponds to, given that these conditions remain present at a certain level in the unexposed. An additional issue, also related to the fact that we are dealing with constructs rather than well-defined interventions, is that we do not know the order of the mediators, which would be needed for confounding control in the simple approach.

We propose the following principles to tackle these issues:

- Explicitly acknowledge that evidence for actual interventions in this context is not possible. Instead, one can address a more modest goal: that of informing "intervention targets", that is, the constructs that future interventions might target, which are what is captured in available data. Although such evidence should be regarded as of lower level than causal inference about well-defined interventions, it might be the only available in the field.

- Define effects that map to a target trial assessing the impact of the distributional mediator shifts that those hypothetical interventions might achieve; these shifts can be individualised, i.e. conditional on covariates. Similar to effects studied by VanderWeele and Hernan,[38] this amounts to setting mediators to random draws from distributions specified to reflect realistic, user-specified benchmarks regarding the potential impacts of hypothetical interventions. The unexposed population (and thus the concept of mediation) regain relevance in specifying these "estimand assumptions". In addition to these, "identification assumptions" are required to ensure that the estimand can be estimated from available data. More assumptions are required for causal inference with ill-defined vs. well-defined interventions, as expected.[10]

- In specifying relevant distributional shifts, consider the joint distribution of the mediators, so that their interrelatedness can be accounted for without causal ordering assumptions. The price to pay for this is a need to make assumptions regarding the correlations amongst the mediators (at a population, distributional level) under the hypothetical interventions, as these correlations cannot be expected to be the same as in the observed data. For example, mental health in a subpopulation offered widespread provision of psychotherapy might be more or less correlated (on average) with substance use than in one offered widespread provision of antidepressants.

Figure 2 provides a conceptual overview of the proposed approach for evaluating hypothetical interventions.

**TARGET TRIAL**

We now describe the target trial that integrates these principles, with focus on three specific policy-relevant questions.

**Question 1: If targeting only one mediator ("one-policy premise"), which of these separate interventions would provide the "biggest bang for the buck"?**

This question is of relevance under resource (e.g. financial) constraints implying that the policy maker would implement only one of the $K$ hypothetical interventions $B_1, \ldots, B_K$, in the exposed population. We make the following estimand assumptions:

E1. Intervention $B_k$ would be applied independently of the other mediators, for $k = 1, \ldots, K$;

E2. Intervention $B_k$ would shift the distribution of mediator $M_k$ to what it would be in the unexposed given $\boldsymbol{C}$, for $k = 1, \ldots, K$. This is equivalent to setting $M_k$ to a random draw from the distribution it would have under no exposure given $\boldsymbol{C}$; and

E3. Intervention $B_k$ would sever the dependence on average between $M_k$ and the other mediators, so that the joint distribution of the other mediators is held at what it would be under exposure, for $k = 1, \ldots, K$.

Formally, we represent E1-E3 as the assumption that the hypothetical intervention $B_k$ would set the mediators to a random draw from the following joint distribution:

$$P(M_{k0} = m_k | \boldsymbol{C}) \times P(\boldsymbol{M}_{(-k)1} = \boldsymbol{m}_{(-k)} | \boldsymbol{C})$$

where $M_{ka}$ denotes the status of $M_k$ when setting $A$ to $a$; $\boldsymbol{M}_{\cdot a}$ denotes the vector $(M_{1a}, \ldots, M_{Ka})$; and $\boldsymbol{M}_{(-k)a}$ denotes $\boldsymbol{M}_{\cdot a}$ without the $k$th component.

Assumption E1 could be modified if the policy maker intended to personalise treatments conditional on other mediators. However, this would require an expanded set of estimand assumptions, e.g. delineating which mediators, etc. Assumption E2 is justified on the basis that realistically we cannot expect effects beyond bringing levels to those in the unexposed, which can be estimated from the data. Other benchmarks could be specified by the user if they make sense in the specific context, but again this may require additional unverifiable assumptions. Assumption E3 can be considered a worst case scenario in the sense that it precludes any effects of the hypothetical intervention flowing onto other mediators that may be causal descendants. We consider that this is suitable for the purpose of comparing potential intervention targets, but it could be relaxed to allow for correlations between $M_k$ and the other mediators under the hypothetical intervention. Although this would be more realistic, it would require further unverifiable assumptions, regarding the extent of such correlations. Indeed, as in the aforementioned psychotherapy versus antidepressant example, the correlation between mediators would not be as in the observed data and rather would depend on the hypothetical intervention. We consider that the reduced set of assumptions (E1-E3) allows for less assumption-laden and thus clearer comparisons and is likely to be widely applicable as a starting point.

A target trial for the self-harm example under these assumptions is depicted in Figure 3. Arms 1 and 2, referred to as the unexposed and exposed groups, correspond to those in a classic two-arm parallel trial design: the intervention is only to set the exposure to $A = 0$ and $A = 1$, respectively, leading to a naturally arising joint distribution of the mediators in each arm. For each of arms 3 to 6, $A$ is set to 1 and in addition one of the hypothetical interventions is applied, shifting the joint distribution of the mediators (given $C$) in some way. For example, in arm 4, intervention $B_2$ is set to 1 so that the distribution of $M_2$ is shifted to be as it is in the unexposed group (following E2) given baseline characteristics but independently of the other mediators (E1), while the joint distribution of $M_1$, $M_3$ and $M_4$ remains as it naturally arises in the exposed group (E3). That is, in arm 4 the intervention regime is to set $(A, B_2)$ to $(1,1)$ and has the effect of setting $A$ to 1, which results in $M_1$, $M_3$ and $M_4$ being set to a random draw from their joint distribution under exposure given $C$; and setting $M_2$ to a random draw from the distribution that it would have had under no exposure given $C$, and this independently from other mediators.

**Question 2: What would be the remaining disparities between exposure groups if it were possible to jointly target all the mediators?**

We can address this question by considering a hypothetical intervention $B_{all}$, targeting all the mediators, and assuming that it shifts the joint distribution of the mediators to be as in the unexposed given $C$ (assumption E2', analogous to E2 for Question 1). Formally, the assumption is that the hypothetical intervention $B_{all}$ sets the mediators to a random draw from the joint distribution $P(\boldsymbol{M}_{.0} = \boldsymbol{m}|C)$.

Arm 7 in Figure 3 shows what this translates to in the target trial: in this arm, $A$ is set to 1 and $B_{all}$ to 1, which shifts the joint distribution of the mediators to what it is in the unexposed group. A different benchmark could be used for E2', but this would require additional assumptions if there are no data from which to estimate it.

**Question 3: What would be the benefit of sequential policies, applying the separate mediator interventions sequentially?**

Let $B_{\{k\}}$ denote an intervention applying all interventions in the sequence $B_1, \dots, B_K$ up to $B_k$ ($k = 1, \dots, K$), so that setting $B_{\{k\}}$ to 1 means that each of $B_1, \dots, B_k$ is set to 1. Interpretation of assumptions E1-E3 is extended to mean that, however it is done (e.g. simultaneously), this results in shifts in the distribution of each mediator $1, \dots, k$ to what it would be in the unexposed given $C$, independently of other mediators and severing the dependence on average from the subsequent ones in the sequence. Formally, the assumption is that $B_{\{k\}}$ is a hypothetical intervention that sets the mediators to a random draw from the joint distribution

$$P(M_{10} = m_1|C) \times \cdots \times P(M_{k0} = m_k|C) \times P(\boldsymbol{M}_{(-1,\dots,-k)1} = \boldsymbol{m}_{(-1,\dots,-k)}|C)$$

with the last factor omitted for $k = K$.

To evaluate the impact of the sequential interventions, we can add arms to the trial, as depicted in Figure 4 for the case of four mediators. Only three arms are added as $B_{\{1\}} = B_1$. For each of arms 8 to 10, $A$ is set to 1 and $B_{\{k\}}$ is set to 1, $k = 2,3,4$. The order of the sequence, here assumed to be $B_1, \dots, B_K$ (E4), should be determined by the research question: which order is of interest from a policy perspective? If a different order were of interest, then the new trial arms, and resulting effects (next section) would be different.

The target trial in Figures 3 and 4 extends in the natural way to the case of $K$ mediators, with $2K + 2$ arms.

**MEDIATION EFFECT DEFINITIONS**

We define interventional effects addressing each question by contrasting the outcome expectation between relevant trial arms. Following the notation in the last column of Figure 3 but considering the general case of $K$ mediators, let $p_{ctr}$ and $p_{trt}$ denote the outcome expectation in the unexposed ("control") and exposed ("treated") groups, respectively; $p_k$, for $k = 1, \dots, K$, denote the outcome expectation in the arm where the distribution of $M_k$ is shifted; and $p_{all}$ the outcome expectation in the arm shifting the joint distribution of all mediators. Further, let $p_{\{0\}} = p_{trt}$ and $p_{\{1\}} = p_1$, and let $p_{\{k\}}$ for $k > 1$ denote the outcome expectation in the arm in which the interventions on $M_1$ to $M_k$ have been applied (Figure 4).

The total causal effect (TCE) in the difference scale is given by: $\text{TCE} = p_{trt} - p_{ctr}$.

**Effects for Question 1: One-policy premise**

We define a type of interventional indirect effect via the $k$th mediator, $\text{IIE}_k$ ($k = 1, \dots, K$), as the contrast between the outcome expectation in the exposed group and the arm in which the $M_k$ distribution is shifted:

$$\text{IIE}_k = p_{trt} - p_k.$$

This quantifies the impact of an intervention targeting $M_k$, while the joint distribution of the other mediators remains as it would be under exposure. In the example, for $M_2$ (weekly cannabis use), $\text{IIE}_2$ is the reduction in risk of financial hardship in self-harmers that would be achieved by reducing their rates of weekly cannabis use to those in the non-self-harmers, while the joint distribution of all other mediators remains unaffected (given covariates). These effects differ from those proposed by Vansteelandt and Daniel[19], which implicitly emulate other distributional shifts.[22]

**Effects for Question 2: Remaining disparities**

We consider the following interventional direct effect not via any mediator (IDE):

$$\text{IDE} = p_{all} - p_{ctr}$$

The IDE quantifies disparities between exposed and unexposed that would remain even if it were possible to intervene simultaneously on all the mediators to shift their joint distribution (mean levels and interdependence) to be as in the unexposed group (given covariates).

**Effects for Question 3: Sequential policies**

We define the interventional indirect effect of the $k$th intervention in the sequence, $\text{IIE}_{\{k\}}$ ($k = 1, \ldots, K$), as:

$$\text{IIE}_{\{k\}} = p_{\{k-1\}} - p_{\{k\}}.$$

The sum of these effects provides an interventional indirect effect quantifying the overall impact of the sequential intervention ($\text{IIE}_{\{\text{seq}\}}$) and is equal to:

$$\text{IIE}_{\{\text{seq}\}} = p_{trt} - p_{\{K\}}$$

**Decompositions of the TCE and other interesting effects**

There are infinite possible decompositions of the TCE and what matters is that the component effects address relevant questions. For example, the TCE is decomposed as: $\text{TCE} = \text{IDE} + \text{IIE}_1 + \cdots + \text{IIE}_K + \text{IIE}_{\text{int}}$, where the last term is a type of interventional indirect effect via the mediators' interdependence, contrasting the benefit of the aforementioned joint intervention with the sum of the benefits of individual interventions: $\text{IIE}_{\text{int}} = (p_{trt} - p_{all}) - (\text{IIE}_1 + \cdots + \text{IIE}_K)$. This effect does not have a policy-relevant interpretation so it is not of much interest.

The decomposition that focusses on sequential policies is: $\text{TCE} = \text{IDE} + \text{IIE}_{\{\text{seq}\}} + \text{IIE}_{\{\text{int}\}}$. Here $\text{IIE}_{\{\text{int}\}}$ contrasts the benefit of the joint intervention $B_{all}$ with the benefit of sequentially applying $B_1, \ldots, B_K$: $\text{IIE}_{\{\text{int}\}} = (p_{trt} - p_{all}) - \text{IIE}_{\{\text{seq}\}} = p_{\{K\}} - p_{all}$. The expression after the second equality shows that this effect better captures what one would intuitively conceive as the effect via the mediators' interdependence: contrasting the expected outcome under a shift in the joint mediator distribution to that when a sequence of independent shifts is made across the mediators.

Other contrasts that could be of interest are $\text{IDE}_k = p_k - p_{ctr} = \text{TCE} - \text{IIE}_k$, for $k = 1, \dots, K$, with $\text{IDE}_k$ quantifying the disparities remaining after intervening on $M_k$ alone. Each effect can be expressed as a proportion of the TCE to gauge relative size.

**IDENTIFICATION AND ESTIMATION**

To identify and simulate these effects it suffices to consider the identifiability and estimation of the outcome expectation in a given target trial arm subject to a mediator distribution shift (arms 3-10). Let $B$ indicate receipt of the corresponding hypothetical intervention (e.g. $B$ stands for $B_1$ in arm 3, $B_{all}$ in arm 7 and $B_{\{2\}}$ in arm 8). Further, for $a, b = 0,1$ and $k = 1, \dots, K$, let: $Y_{ab}$ denote the outcome when $A$ is set to $a$ and $B$ to $b$. Recall that $M_{ka}$ denotes the status of $M_k$ when setting $A$ to $a$; $\boldsymbol{M}_{\cdot a}$ denotes the vector $(M_{1a}, \dots, M_{Ka})$; and $\boldsymbol{M}_{(-k)a}$ denotes $\boldsymbol{M}_{\cdot a}$ without the $k$th component; and denote by $\boldsymbol{M}_{\cdot} = (M_1, \dots, M_K)$ and $\boldsymbol{M}_{(-k)}$, respectively, the observed counterparts of $\boldsymbol{M}_{\cdot a}$ and $\boldsymbol{M}_{(-k)a}$.

In addition to standard positivity assumptions,[39] we make the following identification assumptions:

A1. There is no causal effect of $B$ on the outcome other than through mediator distributional shifts, that is, other than through setting the mediators to a random draw from the specified distribution;

A2. The following conditional independence assumptions hold:

(i) $Y_{ab} \perp (A, B) | \boldsymbol{C}$

(ii) $(M_{1a}, \dots, M_{Ka}) \perp A | \boldsymbol{C}$

A3. $Y_{ab} = Y$ when $A = a$ and $B = b$; $M_{ka} = M_k$ when $A = a$ for $k = 1, \dots, K$

Assumptions A1-A3 are similar to those considered by VanderWeele and Hernan[38]. With the intervention $B$ being hypothetical, it is not possible to assess whether these assumptions are plausible, except for assumptions not pertaining to $B$, which are similar to assumptions in Vansteelandt and Daniel[19]. Further, A3 relies partly on the possibility of identifying the exposure with a well-defined intervention. This can be assessed but, with the main goal being to evaluate mediator interventions, it can be argued that application of the proposed method remains meaningful even with no well-defined exposure intervention, as others have proposed in related settings.[23,24,40]

Under A1-A3, the outcome expectation in the given arm can be emulated using observational data. Complete identification formulae and proofs are given in the Supplementary Materials. For illustration, consider the arm where intervention $B_k$ is applied to shift mediator $k$ under the one-policy premise. From A2(i) and A3, it follows that the outcome expectation $p_k$, can be expressed as: $p_k = E(Y_{11}) = E_C[E(Y|A = 1, B_k = 1, C)]$. By A1, setting $B_k = 1$ is equivalent to setting the mediators to a random draw from the joint distribution $P(M_{k0} = m_k|C) \times P(\boldsymbol{M}_{(-k)1} = \boldsymbol{m}_{(-k)}|C)$, which from A2(ii) and A3 is equal to $P(M_k = m_k|A = 0, C) \times P(\boldsymbol{M}_{(-k)} = \boldsymbol{m}_{(-k)}|A = 1, C)$. This leads to the following identification formula:

$$p_k = E_C\left[\sum_{\boldsymbol{m}=(m_1,\ldots,m_K)} E(Y|A = 1, \boldsymbol{M} = \boldsymbol{m}, C) \times P(M_k = m_k|A = 0, C) \times P(\boldsymbol{M}_{(-k)} = \boldsymbol{m}_{(-k)}|A = 1, C)\right].$$

Estimation can be performed using the Monte Carlo simulation-based g-computation approach described by Vansteelandt and Daniel[19] (see Supplementary Materials). To reduce the risk of misspecification bias, using rich parametric models, including various interaction terms and higher-order terms (for continuous variables), is recommended.[40] Example code in R[41] for implementing the method, including a function and a worked example on simulated data, can be accessed at the first author's GitHub repository (https://github.com/moreno-betancur/medRCT).

**RESULTS FOR SELF-HARM EXAMPLE**

Table 1 shows descriptive statistics based on the 1786 participants (out of 1943 in the cohort study) with the adolescent self-harm exposure available. As all other analysis variables had missing data, subsequent analyses were based on multiple imputation using 40 imputations (details in Supplementary Materials). Table 2 shows preliminary estimates of unadjusted and regression-adjusted exposure-outcome, exposure-mediator and mediator-outcome associations, which were obtained using main-effects multivariable logistic regression models. These provide an idea of the strength of some of the hypothesised pathways in Figure 1.

We estimated the proposed effects using the g-computation method with multivariable logistic regressions including all two-way interactions (see Supplementary Materials). Table 3 shows the results. There was some evidence that adolescent self-harmers had an increased risk of financial hardship in adulthood compared to non-self-harmers: TCE=7.2% (95%CI: -1.7 to 16.1%). Under the one-policy premise, we estimated that the highest impact would be achieved by an intervention that would improve the rates of university completion in adolescent self-harmers (IIE$_3$=0.9%; -1.3 to 3.2%). This corresponds to a 13% reduction in the between-group difference, with the remaining difference being IDE$_3$=TCE − IIE$_3$=6.3%. Other intervention targets have lower impact. A hypothetical intervention shifting the joint distribution of the mediators in the self-harm group to be as under no self-harm, given covariates, would still leave 77% of the difference between the two groups remaining: IDE=5.6% (-3.1 to 14.3%).

The overall sequential policy could, in principle, achieve a reduction of 27% of the total effect (IIE$_{\{seq\}}$=1.9%; -1.4 to 5.2%). This is decomposed into the effects of applying each policy on top of the previous ones in the sequence. Each of the effects from $M_2$ onwards is of slightly lower magnitude than under the one-policy premise. The effect via the interdependence IIE$_{\{int\}}$ is negative, indicating that the sequential intervention would achieve a larger reduction in risk than the joint intervention. This is explained by the severing of dependence amongst the mediators under assumption E3, which, as mentioned, is not a realistic assumption. The direction of this effect indicates that we would estimate a smaller effect for the sequential intervention under the alternative assumption that correlations after the interventions remain as in the observed data (given exposure and confounders).

**DISCUSSION**

While avoiding previous "axiomatic" definitions of mediation, this paper shows that interventional mediation effects provide a vehicle for tackling the issue of ill-defined interventions that abounds in various areas of epidemiology.[10,15–17] Building on previous work[22], novel interventional effects are defined that explicitly emulate target trials of hypothetical interventions that result in individualised (covariate-specific) mediator distributional shifts. Simulating hypothetical interventions in this way addresses the realistic if relatively modest goal of informing intervention targets and requires an expanded set of assumptions both to define the estimand and to identify it with observational data. This is commensurate with the lower-level evidence and increased subtlety in interpretation that is to be expected with ill-defined interventions, towards the left-hand end of the Galea-Hernán causal spectrum,[10] for which one must simulate "in silico hypothetical experiments". Although uncertainty of estimation precludes any strong conclusions being drawn, the self-harm example illustrated the value of our proposal for addressing different policy-relevant questions.

We retained mediation terminology ("direct", "indirect", etc.) for the proposed effects, consistent with the view that there is no clear definition of these notions beyond these and so-called "separable effects" (see below). Interventional direct effects generalise "controlled direct effects", which can be seen as setting the mediator to a draw from a degenerate (constant-valued) distribution. We suggest it is more realistic to focus on the benchmark of our proposed direct effects, the distribution in the unexposed given covariates. Others have also considered more realistic benchmarks in the definition of direct effects.[26,27] More broadly, although the estimand assumptions outlined here are likely to be of relevance in a range of settings, alternative assumptions might well be warranted in other contexts. In particular, further work could consider estimand assumptions that individualise mediator shifts based on a set of baseline covariates that may overlap but is not necessarily equal to the minimal confounding adjustment set $C$.

The identification assumptions that concern hypothetical interventions are not assessable. As has been noted,[10,37,38] confounder selection is complex in this context: considering common causes of the intervention and its target is difficult with no concrete intervention in mind. Nonetheless, the mapping to a target trial makes it clear that all identification assumptions underlying interventional effects would be assessable in randomised experiments of the hypothetical interventions. This contrasts with natural effects, which require empirically unverifiable "cross-world independence" assumptions, [1,6] as well as further untestable assumptions in the context of multiple mediators.[18,22,42,43] This difference is due to interventional effects being population-level quantities, like the total causal effect, whilst natural effects are individual-level effects.[22] An exception for natural effects is when the exposure is separable into components acting through distinct pathways,[6,44,45] with the resulting "separable effects" emulating hypothetical trials of intervention regimes on the exposure components.

Assumptions about the causal ordering of the mediators are not needed for defining and identifying the proposed effects, because estimand assumptions pertain to the joint distribution and, for sequential policies, the choice of question for the policy-maker (e.g. which sequence of policies is of interest?). As previously mentioned, the price to pay for considering the joint distribution in the estimand assumptions is the need for unverifiable assumptions about the dependence between the mediators under the hypothetical interventions, which would differ from that observed in the data. Meanwhile, a (non-causal) ordering needs to be chosen for estimating the joint mediator distribution if a sequential regression approach is used. We implemented g-computation using highly flexible regression models, but parametric misspecification bias is still a possibility.

Applications and future extensions of our proposal, e.g. to time-varying mediators and dynamic policies, should consider the broader set of target trial principles[14]. Development of doubly or multiply robust methods for estimation with machine learning, building on recent work[46], would be desirable to counter parametric misspecification bias. Nonetheless, our proposal opens new avenues for causal inference about policy-relevant effects with ill-defined interventions.

**Software implementation:** Example R code for implementing the method, including a worked example on simulated data, can be accessed at the first author's GitHub repository (https://github.com/moreno-betancur/medRCT).

# REFERENCES

1. Naimi AI, Kaufman JS, MacLehose RF. Mediation misgivings: Ambiguous clinical and public health interpretations of natural direct and indirect effects. *Int J Epidemiol*. 2014;43(5):1656-1661.
2. Robins JM, Greenland S. Identifiability and exchangeability for direct and indirect effects. *Epidemiology*. 1992;3(2):143-155.
3. Pearl J. Direct and indirect effects. In: *Proceedings of the Seventeenth Conference on Uncertainty and Artificial Intelligence.* San Francisco, CA: Morgan Kaufmann; 2001:411–420.
4. VanderWeele TJ. *Explanation in Causal Inference: Methods for Mediation and Interaction*. New York, New York: Oxford University Press; 2015.
5. Imai K, Keele L, Tingley D. Identification, Inference and Sensitivity Analysis for Causal Mediation Effects. *Stat Sci*. 2010;25:51-71.
6. Robins JM, Richardson TS. Alternative graphical causal models and the identification of direct effects. In: Shrout P, Keyes K, Ornstein K, eds. *Causality and Psychopathology: Finding the Determinants of Disorders and Their Cures*. Oxford University Press; 2011:103-158.
7. Hernan MA. Does water kill? A call for less casual causal inferences. *Ann Epidemiol*. 2017;26(10):674-680.
8. Hernan MA. Do you believe in causes? The distinction between causality and causal inference. In: *European Causal Inference Meetings - EuroCIM, Bremen*. ; 2019:1.
9. Galea S. An argument for a consequentialist epidemiology. *Am J Epidemiol*. 2013;178(8):1185-1191.
10. Galea S, Hernán MA. Win-win: Reconciling Social Epidemiology and Causal Inference. *Am J Epidemiol*. 2019.
11. Gelman A, Imbens G. Why ask why? Forward causal inference and reverse causal questions. NBER Working Paper Series, Working Paper 19614. 2013.
12. Holland PW. Causation and Race. *ETS Res Rep Ser*. 2003;2003(1):i-21.
13. Hernán MA, Alonso A, Logan R, et al. Observational studies analyzed like randomized experiments: An application to postmenopausal hormone therapy and coronary heart disease. *Epidemiology*. 2008;19(6):766-779.
14. Hernán MA, Robins JM. Using Big Data to Emulate a Target Trial When a Randomized Trial Is Not Available. *Am J Epidemiol*. 2016;183(8):758-764.

15. Jackson JW, Arah OA. Making Causal Inference More Social and (Social) Epidemiology More Causal. *Am J Epidemiol*. 2020;189(3):179-182.
16. Vanderweele TJ. Counterfactuals in Social Epidemiology: Thinking Outside of "The Box." *Am J Epidemiol*. 2020;189(3):175-178.
17. Robinson WR, Bailey Z. What social epidemiology brings to the table: reconciling social epidemiology and causal inference. *Am J Epidemiol*. 2020;189(3):171-174.
18. VanderWeele TJ, Vansteelandt S, Robins JM. Effect Decomposition in the Presence of an Exposure-Induced Mediator-Outcome Confounder. *Epidemiology*. 2014;25(2):300-306.
19. Vansteelandt S, Daniel RM. Interventional Effects for Mediation Analysis with Multiple Mediators. *Epidemiology*. 2017;28(2):258-265.
20. Geneletti S. Identifying Direct and Indirect Effects in a Non-Counterfactual Framework. *J R Stat Soc Ser B*. 2007;69(2):199-215.
21. Didelez V, Dawid AP, Geneletti S. Direct and Indirect Effects of Sequential Treatments. In: Dechter R, Richardson T, eds. *Proceedings of the 22nd Annual Conference on Uncertainty in Artificial Intelligence*. Arlington, VA: AUAI Press; 2006:138-164.
22. Moreno-Betancur M, Carlin JB. Understanding interventional effects: a more natural approach to mediation analysis? *Epidemiology*. 2018;29(5):614-617.
23. VanderWeele TJ, Robinson WR. On the Causal Interpretation of Race in Regressions Adjusting for Confounding and Mediating Variables. *Epidemiology*. 2014;25(4):473-484.
24. Jackson JW, VanderWeele TJ. Decomposition analysis to identify intervention targets for reducing disparities. *Epidemiology*. 2018;29(6):825-835.
25. Jackson JW, VanderWeele TJ. Intersectional decomposition analysis with differential exposure, effects, and construct. *Soc Sci Med*. 2019;226:254-259.
26. Popham F. Controlled Mediation as a Generalization of Interventional Mediation. *Epidemiology*. 2019;30(3):e21-e22.
27. Naimi AI, Moodie EEM, Auger N, Kaufman JS. Stochastic mediation contrasts in epidemiologic research: Interpregnancy interval and the educational disparity in preterm delivery. *Am J Epidemiol*. 2014;180(4):436-445.
28. Moreno-Betancur M, Koplin JJ, Anne-Louise P, Lynch J, Carlin JB. Measuring the impact of differences in risk factor distributions on cross-population differences in disease occurrence: A causal approach. *Int J Epidemiol*. 2018;47(1):217–225.

29. Hawton K, Saunders KE, O'Connor RC. Self-harm and suicide in adolescents. *Lancet*. 2012;379(9834):2373-2382.

30. Moran P, Coffey C, Romaniuk H, et al. The natural history of self-harm from adolescence to young adulthood: a population-based cohort study. *Lancet*. 2012;379(9812):236-243.

31. Morgan C, Webb RT, Carr MJ, et al. Incidence, clinical management, and mortality risk following self harm among children and adolescents: cohort study in primary care. *BMJ*. 2017;359:j4351.

32. GBD 2016 Disease and Injury Incidence and Prevalence Collaborators, Abajobir AA, Abate KH, et al. Global, regional, and national incidence, prevalence, and years lived with disability for 328 diseases and injuries for 195 countries, 1990-2016: a systematic analysis for the Global Burden of Disease Study 2016. *Lancet*. 2017;390(10100):1211-1259.

33. Bergen H, Hawton K, Waters K, et al. Premature death after self-harm: a multicentre cohort study. *Lancet*. 2012;380(9853):1568-1574. doi:10.1016/S0140-6736(12)61141-6

34. Mars B, Heron J, Crane C, et al. Clinical and social outcomes of adolescent self harm: population based birth cohort study. *BMJ*. 2014;349:g5954.

35. Moran P, Coffey C, Romaniuk H, Degenhardt L, Borschmann R, Patton GC. Substance use in adulthood following adolescent self-harm: A population-based cohort study. *Acta Psychiatr Scand*. 2015;131(1):61-68.

36. Borschmann R, Becker D, Coffey C, et al. 20-year outcomes in adolescents who self-harm: a population-based cohort study. *Lancet Child Adolesc Heal*. 2017;1(3):195-202.

37. Hernan MA, Vanderweele TJ. Compound Treatments and Transportability of Causal Inference. *Epidemiology*. 2011;22(3):368–377.

38. VanderWeele TJ, Hernan MA. Causal inference under multiple versions of treatment. *J Causal Inference*. 2013;1(1):1-20.

39. Hernan MA, Robins J. *Causal Inference: What If*. Boca Raton: Chapman & Hall/CRC; 2020.

40. Micali N, Daniel RM, Ploubidis GB, Stavola BL De. Maternal Prepregnancy Weight Status and Adolescent Eating Disorder Behaviors A Longitudinal Study of Risk Pathways. *Epidemiology*. 2018;29(4):579-589.

41. R Core Team. R: A Language and Environment for Statistical Computing. 2013.

http://www.r-project.org.

42. Daniel RM, De Stavola BL, Cousens SN, Vansteelandt S. Causal mediation analysis with multiple mediators. *Biometrics*. 2015;71(1):1-14.

43. Avin C, Shpitser I, Pearl J. Identifiability of Path-Specific Effects. In: *Proceedings of International Joint Conference on Artificial Intelligence, Edinburgh, Scotland*. ; 2005:357-363.

44. Didelez V. Defining causal meditation with a longitudinal mediator and a survival outcome. *Lifetime Data Anal*. 2019;25(4):593-610.

45. Aalen OO, Stensrud MJ, Didelez V, Daniel R, Røysland K, Strohmaier S. Time-dependent mediators in survival analysis: Modeling direct and indirect effects with the additive hazards model. *Biometrical J*. 2020;62(3):532-549.

46. Benkeser D. *Nonparametric Inference for Interventional Effects with Multiple Mediators*.; 2020. http://arxiv.org/abs/2001.06027. Accessed March 17, 2020.

47. Mars B, Cornish R, Heron J, et al. Using Data Linkage to Investigate Inconsistent Reporting of Self-Harm and Questionnaire Non-Response. *Arch Suicide Res*. 2016;20(2):113-141.

48. Vanderweele TJ. Principles of confounder selection. *Eur J Epidemiol*. 2019.

49. Schomaker M, Heumann C. Bootstrap inference when using multiple imputation. *Stat Med*. 2018;37:2252-2266.

50. van der Laan MJ, Rose S. *Targeted Learning. Causal Inference for Observational and Experimental Data*. New York: Springer; 2011.

51. Van Buuren S, Boshuizen H, Knook D. Multiple imputation of missing blood pressure covariates in survival analysis. *Stat Med*. 1999;18(6):681-694.

52. Tilling K, Williamson EJ, Spratt M, Sterne JAC, Carpenter JR. Appropriate inclusion of interactions was needed to avoid bias in multiple imputation. *J Clin Epidemiol*. 2016;80:107-115.

53. Moreno-Betancur M, Lee KJ, Leacy FP, White IR, Simpson JA, Carlin JB. Canonical Causal Diagrams to Guide the Treatment of Missing Data in Epidemiologic Studies. *Am J Epidemiol*. 2018;187(12):2705-2715.

**FIGURES**

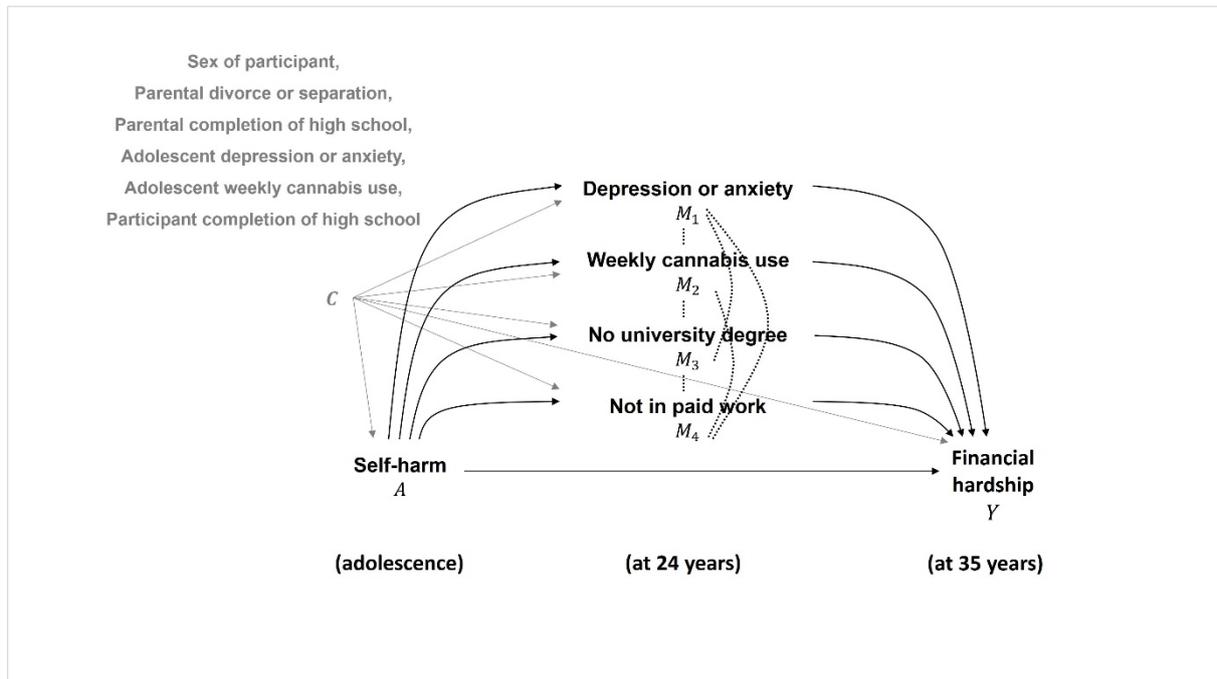

**Figure 1.** Directed acyclic graph portraying the assumed causal structure, conceptualizing the pathways from adolescent self-harm to financial hardship, via the four mediators of interest. Dotted undirected arrows indicate where we are agnostic about the directionality of causal influences. Pre-exposure confounders and arrows from these are shown in grey to improve clarity.

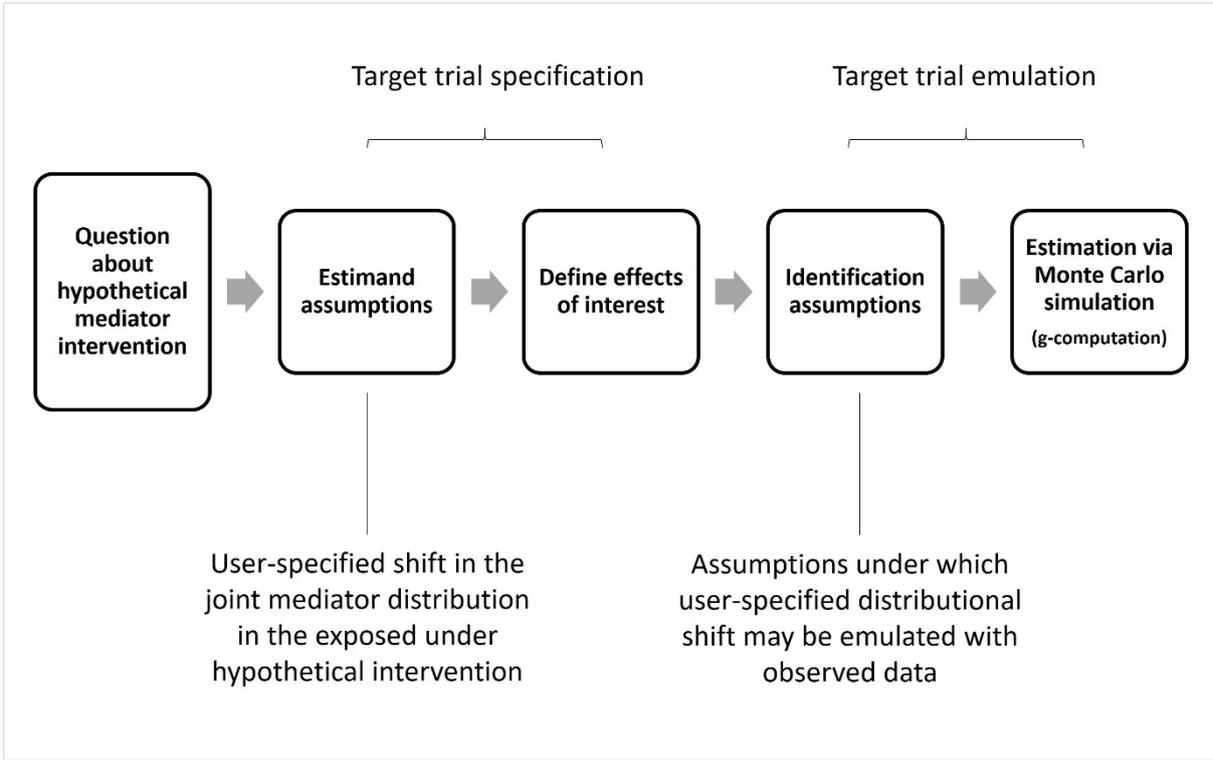

**Figure 2.** Conceptual overview of the proposed approach for tackling the issue of ill-defined interventions via simulation of hypothetical interventions

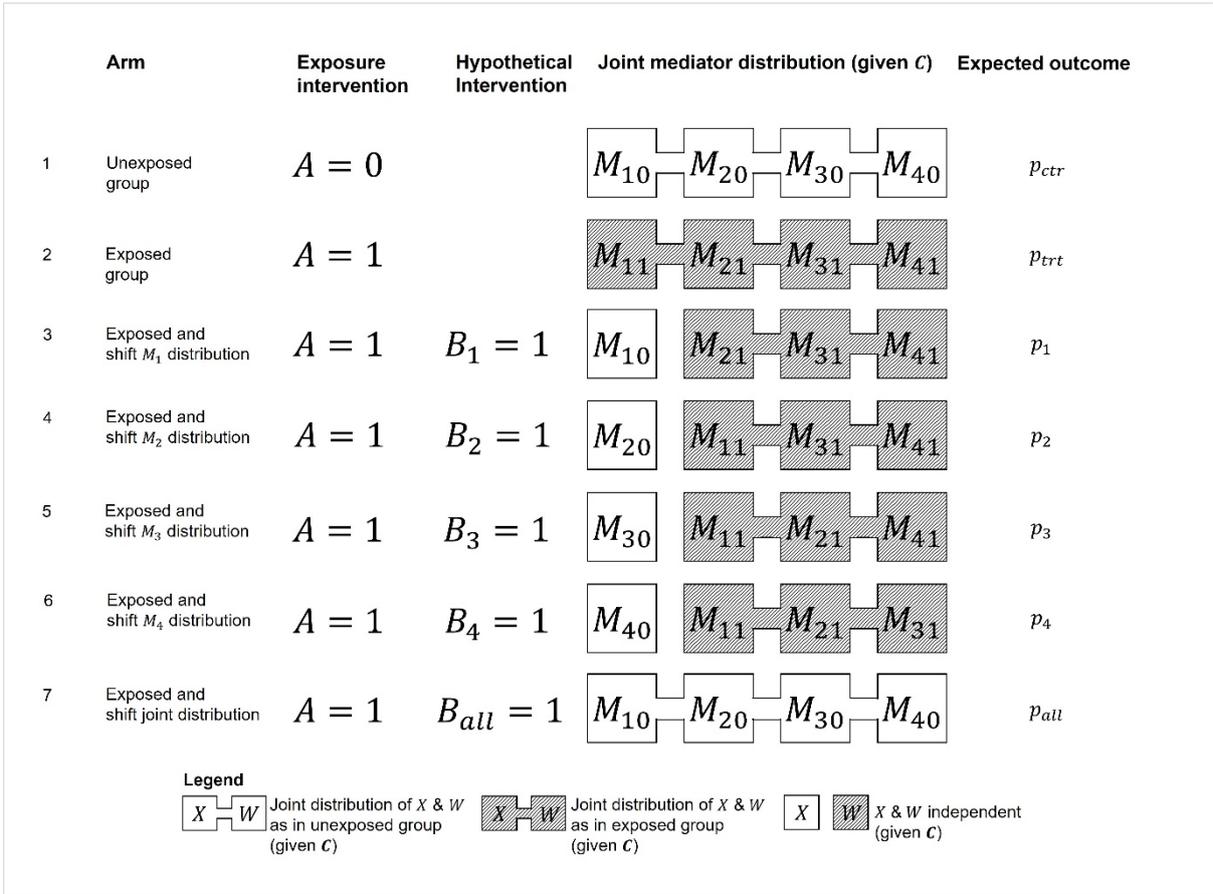

**Figure 3.** Graphical depiction of arms in the "target trial" designed to examine the effects of hypothetical interventions resulting in individualised shifts in the distributions of four interdependent mediators.

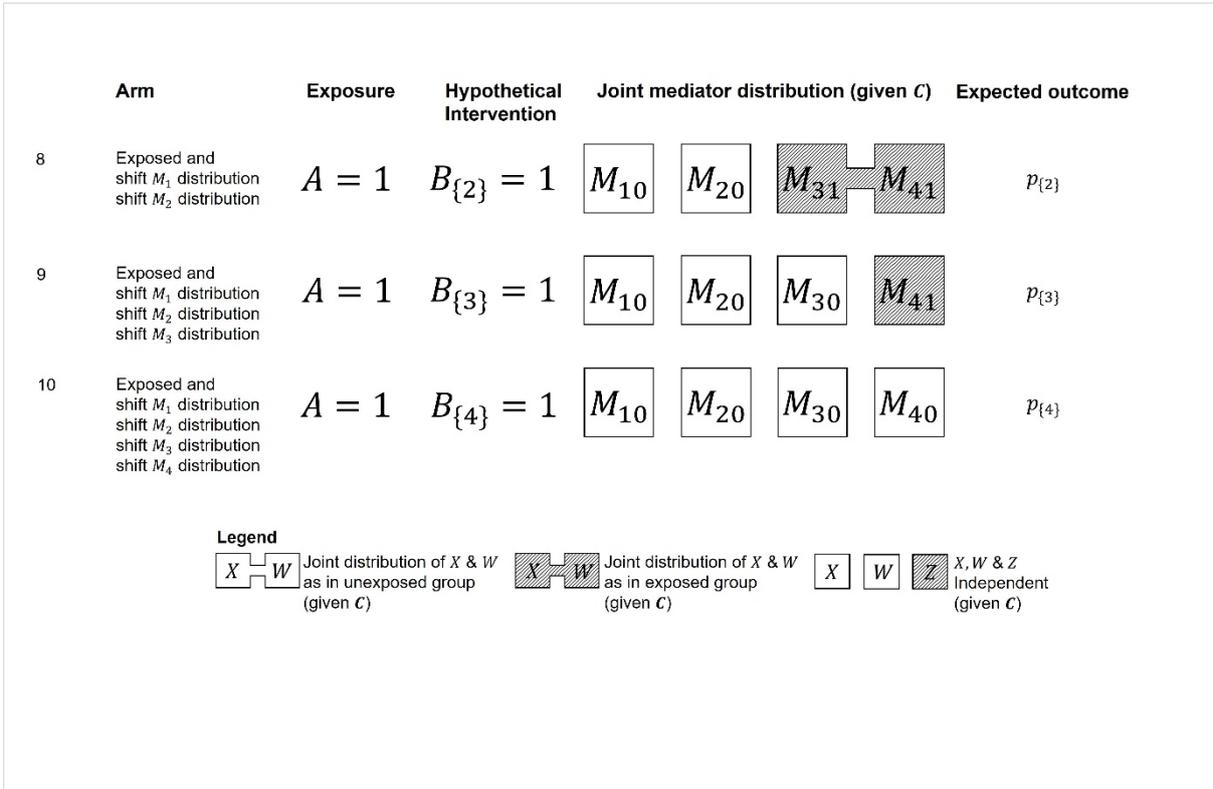

**Figure 4.** Extension of target trial to evaluate the effects of sequentially applying hypothetical interventions resulting in individualised shifts in mediator distributions.

**TABLES**

**Table 1.** Descriptive statistics by exposure group in the self-harm example

|  | Adolescent self-harm[b] | | Missing (%)[c] |
|---|---|---|---|
|  | No | Yes |  |
| Number[a] | 1638 | 148 |  |
| Pre-exposure confounders |  |  |  |
|   Sex of participant: Female (%) | 846 (51.6) | 95 (64.2) | 0.0 |
|   Parental divorce or separation (%) | 339 (20.7) | 45 (30.4) | 0.0 |
|   Neither parent completed secondary school (%) | 515 (32.7) | 46 (33.3) | 4.1 |
|   Adolescent depression or anxiety (%) | 495 (30.2) | 111 (75.0) | 0.0 |
|   Adolescent weekly cannabis use (%) | 155 ( 9.5) | 41 (27.9) | 0.5 |
|   Participant did not complete secondary school (%) | 232 (14.8) | 32 (23.2) | 4.6 |
| Mediators (at age 24 years) |  |  |  |
|   Depression or anxiety (%) | 263 (20.0) | 32 (26.0) | 19.6 |
|   Weekly cannabis use (%) | 143 (10.9) | 25 (20.3) | 19.7 |
|   No university degree (%) | 805 (61.3) | 96 (78.0) | 19.5 |
|   Not in paid work (%) | 140 (10.6) | 22 (17.9) | 19.5 |
| Outcome (at age 35 years) |  |  |  |
|   Financial hardship | 258 (21.9) | 41 (38.3) | 28.0 |
| Any analysis variable missing (%) | 546 (33.3) | 47 (31.8) | 0 |

[a] The total number of participants in each exposure group
[b] Descriptive statistics for each characteristic are based on the records with available data for that variable in the given exposure group
[c] Proportion of missing data across both exposure groups for that variable

**Table 2.** Associations amongst exposure, outcome and mediators estimated using multivariable logistic regression models and multiple imputation (40 imputations)

| Associations | Crude OR | 95% CI | Adjusted OR[a] | 95% CI |
|---|---|---|---|---|
| Exposure (adolescence) - Outcome (35yrs) | | | | |
|     Self-harm - Financial hardship | 2.20 | (1.49; 3.25) | 1.56 | (1.01; 2.42) |
| Exposure (adolescence) - Mediators (24yrs) | | | | |
|     Self-harm - Depression or anxiety | 1.46 | (0.96; 2.22) | 0.93 | (0.59; 1.45) |
|     Self-harm - Weekly cannabis use | 2.06 | (1.31; 3.23) | 1.29 | (0.76; 2.19) |
|     Self-harm - No university degree | 2.07 | (1.34; 3.20) | 1.56 | (0.95; 2.53) |
|     Self-harm - Not in paid work | 1.89 | (1.16; 3.08) | 1.42 | (0.84; 2.40) |
| Mediators (24yrs) - Outcome (35yrs) | | | | |
|     Depression or anxiety - Financial hardship | 1.64 | (1.17; 2.30) | 1.37 | (0.96; 1.95) |
|     Weekly cannabis use - Financial hardship | 1.47 | (1.00; 2.16) | 1.34 | (0.87; 2.08) |
|     No university degree - Financial hardship | 2.97 | (2.16; 4.08) | 2.53 | (1.78; 3.59) |
|     Not in paid work - Financial hardship | 2.23 | (1.53; 3.26) | 1.77 | (1.18; 2.64) |

[a] Adjusted for pre-exposure confounders and, for mediators, the exposure
OR: Odds ratio; CI: Confidence Interval

**Table 3.** Estimates of proposed interventional mediation effects under the one-policy premise and under sequential policies, estimated using the Monte Carlo simulation-based g-computation approach (200 replications), along with the bootstrap (1000 runs) and multiple imputation (40 imputations)

| Effect | | Estimate | 95% CI | Proportion of TCE (%) |
|---|---|---|---|---|
| TCE | | 0.072 | (-0.017; 0.161) | 100 |
| IDE | | 0.056 | (-0.031; 0.143) | 77 |
| **Effects under one-policy premise** | | | | |
| $IIE_1$ | (depression or anxiety) | 0.002 | (-0.015; 0.019) | 3 |
| $IIE_2$ | (weekly cannabis use) | 0.005 | (-0.011; 0.020) | 7 |
| $IIE_3$ | (no university degree) | 0.009 | (-0.013; 0.032) | 13 |
| $IIE_4$ | (not in paid work) | 0.006 | (-0.011; 0.023) | 9 |
| $IIE_{int}$ | (mediators' interdependence) | -0.006 | (-0.021; 0.009) | -8 |
| **Effects under sequential policies** | | | | |
| $IIE_{\{seq\}}$ | (full sequence) | 0.019 | (-0.014; 0.052) | 27 |
| $IIE_{\{1\}}$ | (depression or anxiety) | 0.002 | (-0.015; 0.019) | 3 |
| $IIE_{\{2\}}$ | (weekly cannabis use) | 0.004 | (-0.010; 0.018) | 5 |
| $IIE_{\{3\}}$ | (no university degree) | 0.009 | (-0.012; 0.030) | 12 |
| $IIE_{\{4\}}$ | (not in paid work) | 0.005 | (-0.010; 0.020) | 7 |
| $IIE_{\{int\}}$ | (mediators' interdependence) | -0.003 | (-0.008; 0.002) | -4 |

TCE: Total Causal Effect; IDE: Interventional Direct Effect; IIE: Interventional Indirect Effect; CI: Confidence Interval

# SUPPLEMENTARY MATERIALS

The Supplementary Materials contain further details supplementing the main text of the paper. The contents are as follows:

- Study design, participants and ethics approval in the Victorian Adolescent Health Cohort Study (VAHCS)
- Measures of relevance for the self-harm example in VAHCS
- Description of Monte Carlo simulation-based g-computation estimation method
- Implementation of g-computation method in the self-harm example
- Implementation of multiple imputation in the analysis of the self-harm example
- Identification formulae and proofs
- References for this document

Notation and terminology are as defined in the main text.

**Study design, participants and ethics in the Victorian Adolescent Health Cohort Study**

The Victorian Adolescent Health Cohort Study is a 10-wave longitudinal cohort study of health across adolescence to the fourth decade of life in the state of Victoria, Australia, conducted between August 1992 and March 2014. At baseline, a representative sample of mid-secondary school adolescents was selected with a two-stage cluster sampling procedure. At stage one, 45 schools were chosen at random from a stratified frame of government, Catholic, and independent schools, with a probability proportional to the number of Year 9 (aged 14–15 years) students in the schools in each stratum. At stage two, one single intact class was selected at random from each participating school in the latter part of the ninth school year (wave 1), and a second class from each school was selected 6 months later (wave 2). One school did not continue beyond wave 1, causing a loss of 13 participants and leaving 44 schools in the study. Participants were reviewed at four 6-month intervals between the ages of 15–18 years (waves 3–6) with four follow-up waves in adulthood, ages 20–21 years (wave 7), 24–25 years (wave 8), 28–29 years (wave 9), and 34–35 years (wave 10).

Data collection protocols were approved by the Ethics in Human Research Committee of the Royal Children's Hospital, Melbourne. Informed parental consent was obtained before inclusion in the study. In the adult phase, all participants were informed of the study in writing and gave verbal consent before being interviewed.

**Measures in the Victorian Adolescent Health Cohort Study**

*Exposure (A):* Adolescent self-harm was assessed at each of waves 3 to 6 using the following question: "In the last [reference period] have you ever deliberately hurt yourself or done anything that you knew might have harmed you or even killed you?" The reference period was 1 year for wave 3 and 6 months for waves 4 to 6. Participants who responded positively to the main question were then asked to describe the nature and timing of each self-harm event. These detailed responses were coded into five subtypes of self-harm (by the study's principal investigator, confirmed by a co-investigator): cutting or burning, self-poisoning, deliberate and potentially life-threatening risk-taking, self-battery, and other (including attempted self-drowning, hanging, intentional electrocution and suffocating). Participants were classified at each wave with "any self-harm" if they were identified to have reported one or more of these individual categories.

A summary measure across adolescence was used in analyses, defined as any occurrence in waves 3–6, with a negative value assumed when all wave-specific measures were either negative or missing. We define $A = 1$ if self-harm was present according to this summary measure and $A = 0$ otherwise. Our choice to collapse the self-harm exposure over the adolescent period in this way is justified on the basis that this provided a more robust measure of this rare event over this period, in particular helping us capture more cases/reduce measurement error given that this is self-reported. Indeed, it is has been found that self-harm events tend to be under-ascertained.[47]

*Outcome (Y):* At age 35 years (wave 10), the outcome measure of financial hardship was assessed via a positive response to one or more of the following: ["Over the past 12 months, due to a shortage of money, you.."] 1) have been unable to pay gas, electricity, or telephone bills on time; 2) have been unable to pay mortgage or rent on time; 3) could not afford a night out once a fortnight; and/or 4) could not afford a holiday away for at least 1 week a year. We define $Y = 1$ if financial hardship was present and $Y = 0$ otherwise.

*Mediators ($M_1, M_2, M_3, M_4$):* Potential mediating factors, measured at age 24 (wave 8), were depression and/or anxiety ($M_1$), weekly or more frequent cannabis use over the past year ($M_2$), not having completed a university degree ($M_3$), and not being in paid work ($M_4$). At wave 8, symptoms of depression and anxiety were assessed with the 12-item General Health Questionnaire ("GHQ-12"), an extensively used screening tool for psychological disorders in general health care, dichotomised at the cut-off point of 3 or above to indicate a level of distress appropriate for clinical intervention. We defined $M_k = 1$ if the mediator is present and $M_k = 0$ if it is absent.

*Pre-exposure confounders (C)*: These were selected on an a priori basis considering potential confounders of the exposure-mediator, mediator-outcome and exposure-outcome associations. We selected: participant sex, parental completion of high school, parental divorce or separation up to and including wave 6, and adolescent antecedents of the mediators where present, specifically participant completion of high school, adolescent depression and/or anxiety and cannabis use (weekly or more frequent). The latter two were summarised across waves 3-6 in the same way as the exposure. Symptoms of depression and anxiety were assessed using the revised Clinical Interview Schedule ("CIS-R"). The total scores on this scale were dichotomised at a cut-off point of 11 (≤11 *vs* >11) to delineate a mixed depression-anxiety state, at a lower threshold than syndromes of major depression and anxiety disorder but for which clinical intervention would still be appropriate.

Of note, we assume that the confounders (**C**) precede the exposure, even though some are measured contemporaneously with it, specifically: parental divorce or separation by end of wave 6, adolescent depression or anxiety (waves 3-6), adolescent weekly cannabis use (waves 3-6) and participant completion of high school (approximately concurrent with wave 6). This choice follows a principle proposed by VanderWeele[48] to deal with the challenging issue of confounder selection in real-world studies where knowledge regarding the causal diagram is limited, as in our example. Specifically, he suggests to include proxies for unmeasured variables that are common causes of both the exposure and the outcome. We consider that the aforementioned measures are proxies of pre-exposure confounders (e.g. low school engagement or family difficulties over high school) and thus by adjusting for them we hope to have limited the extent of residual confounding.

**Description of estimation method**

As mentioned in the main text, estimation of the proposed effects can be conducted using a Monte Carlo simulation-based g-computation approach as described in Vansteelandt and Daniel.[19] It relies on the factorization of joint mediator distributions as the sequential product of conditional distributions. The procedure is described next and in each step we exemplify what it entails in the context of estimating the second term of $IIE_2$, $p_2$, which is identified by:

$E_C[\sum_{m=(m_1,\ldots,m_K)} E(Y|A = 1, M = m, C) \times P(M_2 = m_2|A = 0, C) \times P(M_{(-2)} = m_{(-2)}|A = 1, C)]$.

Step 1. Fit regression models to the observed data to estimate the required distributions. For $p_2$ these are $P(M_2 = m_2|A, C)$, $P(M_1 = m_1|A, C)$, $P(M_3 = m_3|M_1, A, C)$, ..., $P(M_K = m_K|M_1, M_3, \ldots, M_{K-1}, A, C)$ and $E(Y|A, M_1, \ldots, M_K, C)$. All these models should include as many higher-order interactions amongst exposure and mediators as supported by the data.

Step 2. For each individual, sequentially draw mediator values from the fitted distributions, given their covariate vector $c$, the relevant exposure value (in the example, $A = 0$ for $M_2$, $A = 1$ for other mediators) and draws of previous mediators as required (in the example, required from $M_3$ onwards).

Step 3. Using the fitted outcome model, for each individual predict the outcome given the value of their covariate vector $C$, the relevant exposure value (in the example, $A = 1$) and the mediator draws.

Step 4. Repeat steps 1 to 3 a large number of times and average the outcome predictions across the whole sample and these repetitions to obtain an estimate of the required term.

Steps 1 to 4 are a way of estimating a weighted average, where the weights reflect the required population-level intervention on the joint distribution of the mediators.[22] The estimated terms are then added or subtracted to obtain the required contrast i.e. effect estimate. The nonparametric bootstrap can be used to obtain standard errors, confidence intervals and p-values. Example code in R for implementing the method, including a function and a worked example on simulated data, is provided (details provided in the title page only to protect blinding).

To reduce the risk of misspecification bias, using rich parametric models, including various interaction terms and higher-order terms, is recommended.[40] Vansteelandt and Daniel[19] acknowledge that it is possible that the models used in the estimation procedure are not compatible with each other, for example the two different models used for $M_3$ in estimating $p_2$ and $p_3$, but note that this not likely to be more problematic in practice than the overall issue of misspecification bias. Finally, we note that estimation as above, using factorizations of joint distributions as sequential products, requires choosing a (non-causal) ordering and different choices might lead to different estimates. This is part of the broader misspecification bias concern.

**Implementation of g-computation method in the self-harm example**

We estimated the proposed effects using the g-computation method with multivariable logistic regressions. To reduce the risk of misspecification, in our example we followed a strategy similar to that of Micali et al,[40] making the parametric models used in the g-computation procedure rich and flexible by including interaction terms. We note that our sample size was smaller than in that study, and also note that all our variables were binary meaning that we had no concerns regarding violation of linearity assumptions as in that paper. Specifically, to build our models, we progressively included all two-way interactions then all three-way amongst exposure and mediators in all models (for mediators and outcome). Only two-way interactions were supported by the data in all models, so we used this specification. We felt confident about our results when observing that the relative strength of the mediation effects estimated were in accordance with the relative strengths of exposure-mediator and mediator-outcome associations observed in the preliminary analyses based on single parametric models (Table 2 in main text).

We conducted 200 simulations and standard errors were obtained using 1000 nonparametric bootstrap samples under a "multiple imputation then bootstrap" approach. The latter is theoretically valid when estimators are normally distributed,[49] which is the case for g-computation estimators based on maximum likelihood under the assumption that the parametric models are correctly specified.[50]

**Implementation of multiple implementation in self-harm example**

A multiple imputation approach, described next, was used to handle missing data for both the preliminary analyses of associations and the mediation analyses in the self-harm example.

Multiple imputation by chained equations[51] was used with 40 imputations and a logistic regression imputation model for each variable including all analysis variables, three auxiliary background variables associated in the sample with incomplete participation (school region on entry to study, at least one parent smokes cigarettes most days, no parent drinks alcohol most days), relevant auxiliary variables from the preceding wave (e.g. wave 7 mental health to impute wave 8 mental health), and all relevant interactions for the mediation analysis models as recommended.[52] The inclusion of auxiliary variables was intended to make the "missing at random" assumption more plausible, although we note that the missing at random assumption is a sufficient but not a necessary assumption for approximately unbiased estimation with multiple imputation.[53]

**Identification formulae and proofs**

To identify each of the effects defined it suffices to consider the identifiability of the outcome expectation in each arm of the target trial. This is because all effects are contrasts of these parameters. Identifiability formulae and proofs for the outcome expectations $p_{ctr}$ and $p_{trt}$ in arms where only the main exposure is intervened upon (arms 1 and 2 in Figure 3 of main text) are well known in the literature.[39] Thus we next focus on identifiability results for the outcome expectation in arms subject to a hypothetical intervention resulting in mediator distributional shifts (arms 3-10 in Figure 3 of main text). We first recapitulate and extend the notation, and also recall the assumptions as in the main text.

*Notation*

Following notation in the main text, for a given target trial arm, let $A$ be the main exposure and $B$ indicate receipt of the corresponding hypothetical intervention (e.g. $B$ stands for $B_1$ in arm 3, for $B_{all}$ in arm 7 and $B_{\{2\}}$ in arm 8). Further, for $a, b = 0,1$ and $k = 1, \ldots, K$ let: $Y_{ab}$ or $Y_{A=a,B=b}$ denote the outcome when $A$ is set to $a$ and $B$ to $b$; $M_{ka}$ denote the status of mediator $M_k$ when setting $A$ to $a$; $\boldsymbol{M}_{\cdot a}$ denote the vector $(M_{1a}, \ldots, M_{Ka})$; $\boldsymbol{M}_{(-k)a}$ denote $\boldsymbol{M}_{\cdot a}$ without the $k$th component.. Denote by $\boldsymbol{M}. = (M_1, \ldots, M_K)$ and $\boldsymbol{M}_{(-k)}$ respectively, the observed counterparts of $\boldsymbol{M}_{\cdot a}$ and $\boldsymbol{M}_{(-k)a}$. We extend the notation as follows: let $\boldsymbol{M}_{(-1,\ldots-k)a}$ denote $\boldsymbol{M}_{\cdot a}$ without components 1 through to $k$, where $\boldsymbol{M}_{(-1,\ldots,-K)a}$ is null, and $\boldsymbol{M}_{(-1,\ldots,-K)a}$ is the observed counterpart.

*Assumptions*

In addition to standard positivity assumptions, we assume A1-A3 in the main text, which are:

A1. There is no causal effect of $B$ on the outcome other than through mediator distributional shifts, that is, other than through setting the mediators to a random draw from the specified distribution;

A2. The following conditional independence assumptions hold:

(iii) $\quad Y_{ab} \perp (A, B) | \boldsymbol{C}$

(iv) $\quad (M_{1a}, \ldots, M_{Ka}) \perp A | \boldsymbol{C}$

A3. $Y_{ab} = Y$ when $A = a$ and $B = b$; $M_{ka} = M_k$ when $A = a$ for $k = 1, \ldots, K$

*Identification formulae*

Under the above assumptions, the outcome expectations in the target trial arms can be expressed in terms of observed data as follows:

For $k = 1, \ldots, K$,

$$p_k = E_{\boldsymbol{C}} \left[ \sum_{\boldsymbol{m}=(m_1,\ldots,m_4)} E(Y|A=1, \boldsymbol{M}=\boldsymbol{m}, \boldsymbol{C}) \times P(M_k = m_k | A = 0, \boldsymbol{C}) \times P(\boldsymbol{M}_{(-k)} = \boldsymbol{m}_{(-k)} | A = 1, \boldsymbol{C}) \right]$$

$$p_{all} = E_{\boldsymbol{C}} \left[ \sum_{\boldsymbol{m}=(m_1,\ldots,m_4)} E(Y|A=1, \boldsymbol{M}=\boldsymbol{m}, \boldsymbol{C}) \times P(\boldsymbol{M} = \boldsymbol{m} | A = 0, \boldsymbol{C}) \right]$$

$$p_{\{k\}} = E_C\left[\sum_{m=(m_1,\ldots,m_4)} E(Y|A=1, M=m, C) \times P(M_1 = m_1|A=0, C) \times \cdots \times P(M_k = m_k|A=0, C)\right.$$
$$\left. \times P(M_{(-1,\ldots,-k)} = m_{(-1,\ldots,-k)}|A=1, C)\right]$$

where the last factor in the last expression is omitted for $k = K$.

*Proofs*

These proofs closely follow those in Vansteelandt and Daniel[19].

To proof the first identification formula, note that for $k = 1, \ldots, K$, we have:

$$p_k = E(Y_{A=1, B_k=1})$$

$$= E_C[E(Y|A=1, B_k=1, C)]$$

$$= E_C\left[\sum_{m=(m_1,\ldots,m_K)} E(Y|A=1, M=m, C) \times P(M_{k0} = m_k|C) \times P(M_{(-k)1} = m_{(-k)}|C)\right]$$

$$= E_C\left[\sum_{m=(m_1,\ldots,m_4)} E(Y|A=1, M=m, C) \times P(M_k = m_k|A=0, C) \times P(M_{(-k)} = m_{(-k)}|A=1, C)\right]$$

The first equality is by definition; the second equality follows from the law of total probability, conditional independence property A2(i) and the consistency assumption A3; the third equality follows from A1, which means that setting $B_k = 1$ is equivalent to setting the mediators to a random draw from the joint distribution $P(M_{k0} = m_k|C) \times P(M_{(-k)1} = m_{(-k)}|C)$; the fourth equality follows from A2(ii) and the consistency assumption A3.

As for the second identification formula,

$$p_{all} = E(Y_{A=1, B_{all}=1})$$

$$= E_C[E(Y|A=1, B_{all}=1, C)]$$

$$= E_C\left[\sum_{m=(m_1,\ldots,m_K)} E(Y|A=1, M=m, C) \times P(M_{\cdot 0} = m|C)\right]$$

$$= E_C\left[\sum_{m=(m_1,\ldots,m_4)} E(Y|A=1, M=m, C) \times P(M_\cdot = m|A=0, C=c)\right]$$

The first equality is by definition; the second equality follows from the law of total probability, conditional independence property A2(i) and the consistency assumption A3; the third equality follows from A1, which means that setting $B_{all}=1$ is equivalent to setting the mediators to a random draw from the joint distribution $P(M_{\cdot 0} = m|C)$; the fourth equality follows from A2(ii) and the consistency assumption A3.

Finally, for the third identification formula, we have for $k=1,\ldots,K,$:

$$p_{\{k\}} = E\left(Y_{A=1, B_{\{k\}}=1}\right)$$

$$= E_C[E(Y|A=1, B_{\{k\}}=1, C)]$$

$$= E_C\left[\sum_{m=(m_1,\ldots,m_K)} E(Y|A=1, M=m, C) \times P(M_{10} = m_1|C) \times \cdots \times P(M_{k0} = m_k|C) \right.$$
$$\left. \times P(M_{(-1,\ldots,-k)1} = m_{(-1,\ldots,-k)}|C)\right]$$

$$= E_C\left[\sum_{m=(m_1,\ldots,m_4)} E(Y|A=1, M=m, C) \times P(M_1 = m_1|A=0, C) \times \cdots \times P(M_k = m_k|A=0, C) \right.$$
$$\left. \times P(M_{(-1,\ldots,-k)} = m_{(-1,\ldots,-k)}|A=1, C)\right]$$

The first equality is by definition; the second equality follows from the law of total probability, conditional independence property A2(i) and the consistency assumption A3; the third equality follows from A1, which means that setting $B_{\{k\}} = 1$ is equivalent to setting the mediators to a random draw from the joint distribution $P(M_{10} = m_1|C) \times \cdots \times P(M_{k0} = m_k|C) \times P(\boldsymbol{M}_{(-1,\ldots,-k)1} = \boldsymbol{m}_{(-1,\ldots,-k)}|C)$, with the last factor omitted for $k = K$; the fourth equality follows from A2(ii) and the consistency assumption A3.